\documentclass[aps,prb,twocolumn,superscriptaddress,showpacs,floatfix]{revtex4}

\usepackage{graphicx}

\begin{document}

\title{Effect of pinning on the vortex-lattice melting line in
type-II superconductors}

\author{G.~P.~Mikitik}
\affiliation{Max-Planck-Institut f\"ur Metallforschung,
   D-70506 Stuttgart, Germany}
\affiliation{B.~Verkin Institute for Low Temperature Physics
   \& Engineering, Ukrainian Academy of Sciences,
   Kharkov 61103, Ukraine}
\author{E.~H.~Brandt}
\affiliation{Max-Planck-Institut f\"ur Metallforschung,
   D-70506 Stuttgart, Germany}

\date{\today}

\begin{abstract}

The vortex-lattice melting line in three-dimensional type-II
superconductors with pinning is derived by equating the free
energies of the vortex system in the solid and liquid phases.
We account for the elastic and pinning energies and the entropy
change that originates from the disappearance of the phonon
shear modes in the liquid. The pinning
is assumed to be caused by point defects and to be not too strong
so that the melting line lies inside the so-called bundle-pinning
region. We show that the derived equation for the melting line is
equivalent to some Lindemann criterion, which however differs from
that used previously. Estimating the effect of pinning on the
entropy jump at melting, we find the upper critical point of the
melting line from the condition that this jump vanishes.
We also consider the $H$-$T$ phase diagrams of type-II
superconductors for different strengths and types of pinning and
analyze the two recently discussed scenarios how the melting line
and the order--disorder line merge.
\end{abstract}


\maketitle

\section{Introduction}  

\begin{figure}
\includegraphics[scale=.5]{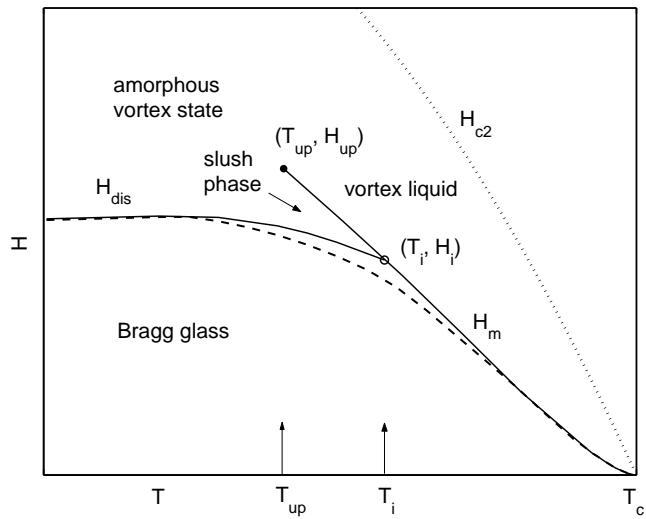}
\caption{\label{fig1} Schematic $T$-$H$ phase diagrams for the
first (solid lines) and for the second (dashed line) scenarios. In
the first scenario the melting line terminates at the so-called
upper critical point ($T_{\rm up}$, $H_{\rm up}$) which in general
does not coincide with the intersection point ($T_i$, $H_i$). In
this case a slush phase (i.e., a vortex liquid with smaller
density of dislocations) can be observed. In the second scenario
the order--disorder and the melting lines are manifestations of
a unified phase transition line. Note that for both scenarios
the vortex liquid and the amorphous vortex state are, in fact, one
and the same phase, which has different viscosity at low and high
temperatures. (We do not discuss the ``vortex glass transition''
which may be not a true phase transition.)
 } \end{figure}

In three-dimensional high-$T_c$ superconductors with pinning, two
phase transition lines are known to exist in the magnetic field
$H$ - temperature $T$ plane:\cite{1,2,3,4,5} The line $H_m(T)$
where a quasiordered Bragg glass\cite{6,7} thermally melts into a
flux-line liquid, and the order--disorder transition line
$H_{dis}(T)$ separating the Bragg glass from an amorphous vortex
state. The melting is caused by thermal vibrations of the lattice,
while the order--disorder transition is induced by quenched
disorder in the vortex system. These two lines merge at some point
in the $H$-$T$ plane. Although both transitions are accompanied by
a proliferation of dislocations in the vortex lattice, it was
argued \cite{8} that the dislocation density $\rho$ is essentially
different in these cases: $\rho \sim a^{-2}$ for melting, and
$\rho \sim R_a^{-2}$ for the order--disorder transition. Here $a$
is the spacing between flux lines, and $R_a$ is the so-called
positional correlation length \cite{9} within which the relative
vortex displacements caused by the quenched disorder are of the
order of $a$. In fact, an intersection of these two different
phase transition lines occurs in this scenario, and the
order--disorder line terminates at the intersection point while
the melting line continues for some distance to higher $H$, see
Fig.~\ref{fig1}. Within this physical picture, the existence of
the so-called slush phase\cite{10} can be naturally explained.
Recent experiments \cite{11,12,13,14,15} for YBaCuO seem to
support this scenario. On the other hand, experimental data
\cite{16,17,18} for BSCCO strongly argue in favor of a different
scenario which was implied, e.g., in the Refs.~\onlinecite{2,5}.
In this second scenario, the dislocation densities for both lines
coincide at the point where these lines merge, and in fact, one
deals with only one phase transition line that describes both the
order--disorder transition at low temperature and the melting near
$T_c$, Fig.~\ref{fig1}.

Phase diagrams of superconductors with pinning reflect the
competition of three characteristic energies \cite{3}: the elastic
energy, the pinning energy, and the energy of thermal
fluctuations. At melting, the cost in the elastic energy due to
the proliferation of dislocations is mainly balanced by the
entropy gain associated with thermal fluctuations, while the role
of the pinning energy, as shown below, is determined
by the parameter $a/R_c$. Here $R_c$ is the transverse collective
pinning length \cite{9}. On the other hand, at the order--disorder
transition, the balance of pinning energy and elastic energy is
most important, while the relative contribution of the entropy
gain is negligible at low temperatures and, according to the
scenario of Ref.~\onlinecite{8}, is determined by the ratio
$a/R_a$ near the intersection point. Thus, if the intersection of
the melting and the order--disorder lines occurred sufficiently
deep in the bundle pinning region (so that $R_c \gg a$ at this
point), the scenario of Ref.~\onlinecite{8} would lead to the
conclusion that up to the intersection point, one can find the
melting line by neglecting the pinning, and the order--disorder
line by neglecting the entropy gain. Just this approximation was
used in our paper\cite{4} for analyzing the phase diagrams of
superconductors. However, our recent results\cite{19} point out
that flux-line pinning can affect the melting line near the
intersection point since at this point, the ratio $R_c/a$ has
decreased to several units for any magnitude of the quenched
disorder in the vortex lattice (even when the disorder is weak).
As to the second scenario, the three energies are all the same
order of magnitude in the temperature region where the
order--disorder transition gradually transforms into melting.
Thus, whatever scenario occurs in reality, a detailed
investigation of the effect of pinning on the melting line is
important to clarify the most intriguing part of the phase
diagram.

The effect of pinning by point defects on the melting line was
observed both in BSCCO\cite{20}and in YBaCuO \cite{21,22,23}
crystals. It was discovered that an increase of the quenched
disorder in the vortex lattice leads to a noticeable shift of the
intersection point to lower magnetic fields and simultaneously
pushes the melting line in the $T$-$H$ plane slightly {\it
downwards}, i.e., at a fixed temperature the appropriate magnetic
field of the melting, $H_m(T)$, decreases. It is important that
the shift of the intersection point is essentially more pronounced
than the decrease of $H_m(t)$ itself.

Some theoretical results on this subject were obtained in
Refs.~\onlinecite{24,2,5}. Larkin and Vinokur \cite{24} started
from the assumption that for the vortex lattice with quenched
disorder to melt, the temperature must match a characteristic
barrier composed of the elastic energy, $E_{\rm el}$, and the
pinning energy, $E_{\rm pin}$. So they estimated the effect of
pinning on the melting line by considering the following balance
of these three energies:
\begin{equation}\label{1}
 T=E_{\rm pin}+E_{\rm el},
\end{equation}
where $E_{\rm pin}$  and $E_{\rm el}$ were calculated in the
so-called cage model.\cite{1} But it follows from this equation
that the melting line has to shift {\it upwards} when the quenched
disorder increases. Another approach was used in
Ref.~\onlinecite{2}. To describe the unified phase transition
line, Giamarchi and Le Doussal, \cite{2} who implied the second
scenario, put forward the following generalization of the
Lindemann criterion:
\begin{equation}\label{2}
  u_{\rm total}^2=c_L^2a^2,
\end{equation}
where $c_L$ is the Lindemann constant ($c_L\sim 0.1-0.2$), and
$u_{\rm total}\sim [2u_T^2+u^2(a,0)]^{1/2}$ is the rms
displacement of neighboring flux lines caused both by the thermal
fluctuations and by the quenched disorder in the lattice. Here
$u_T$ is the magnitude of the thermal fluctuations, while $u(a,0)$
describes the mean relative displacement of neighboring flux lines
caused by the disorder. This criterion leads to the usual
condition for the order--disorder transition \cite{1,3,4} at low
temperatures when $u_T$ is negligible, and it goes over to the
well known Lindemann criterion for pure melting when the disorder
disappears. Equation (\ref{2}) results in a qualitatively correct
dependence of the melting line on pinning by point defects.
However, this reasonable criterion is only an interpolation
formula between the two limiting cases and has no serious
justification. The same is true for the criterion of
Ref.~\onlinecite{5}. Radzyner et al.\cite{5} described the unified
phase transition line using the criterion,
\begin{equation}\label{3}
 u_T^2+ u^2(a,0)=c_L^2a^2,
\end{equation}
which practically coincides with Eq.~(\ref{2}). The criterion
(\ref{3}) is equivalent to the following balance of energies:
\begin{equation}\label{4}
 T+E_{\rm pin}=E_{\rm el},
\end{equation}
which evidently differs from Eq.~(\ref{1}). In calculations of the
phase transition line, the energies $E_{\rm pin}$ and $E_{\rm el}$
were estimated in Ref.~\onlinecite{5} in the framework of the cage
model. But it remained unclear why the pinning energy $E_{\rm
pin}$ now enters into Eq.~(\ref{4}) with the opposite sign as
compared to the energy balance (\ref{2}) suggested by Larkin and
Vinokur. \cite{24}

In the present paper, in order to find and to justify a criterion
for the melting of the flux-line lattice with quenched disorder,
we start with the analysis of  melting in the ideal lattice and
show that three different approaches lead to the same dependence
$H_m(T)$. These approaches are: the Lindemann criterion, the
energy balance, and the rigorous approach based on the
Ginzburg-Landau (or on the London) Hamiltonian. In the case of the
lattice with quenched disorder, we show that the pinning energy in
the flux-line liquid is {\it larger} than the pinning energy in
the Bragg glass. For this reason, the difference of these pinning
energies, $E_{\rm pin}$, has the opposite sign as compared to
Eq.~(\ref{1}), and we arrive at an equation similar to
Eq.~(\ref{4}) but with an expression for $E_{\rm pin}$ that
differs from the estimates previously published. Besides this, we
estimate the effect of pinning on the entropy gain at melting and
find the upper critical point of the melting line from the
condition that this gain vanishes. We also show that the result for
$H_m(T)$ based on this energy balance agrees with the result which
can be derived from the Ginzburg-Landau Hamiltonian.
Then, using the $H_m(T)$ obtained in the framework of the second
and the third approaches, we find how the Lindemann criterion
should be modified to give the same melting line. Interestingly,
the presented energy balance clarifies the difference between the
first and the second scenarios. Finally, we present the $T$ - $H$
phase diagrams of superconductors with two types of flux-line
pinning by point defects and compare these diagrams for the two
scenarios.

In this paper we consider only magnetic fields exceeding
considerably the lower critical field $H_{c1}$ and thus do not
distinguish between the magnetic field $H$ and the magnetic
induction $B$. Besides this, we deal only with uniaxial
anisotropic three-dimensional superconductors, neglecting
completely the decoupling of the superconducting layers. The
anisotropy is characterized by the parameter
$\epsilon=\lambda_{ab}/\lambda_c <1$ where $\lambda_{ab}$ and
$\lambda_c$ are the London penetration depth in the plane $ab$
perpendicular to the anisotropy axis and along this axis,
respectively. The magnetic field is assumed to be directed along
the anisotropy axis. As to the  quenched disorder in the flux-line
lattice, we assume that it is caused by point defects and is
not too strong such that the melting line lies entirely in the
bundle pinning region.

 \section{melting of the ideal vortex lattice} 

We begin with the analysis of melting in the ideal pinning-free
vortex lattice and compare the results of various approaches.

\subsection{Lindemann criterion}  

According to the well known Lindemann criterion, the flux-line
lattice melts when the magnitude of the thermal displacements of
the lattice relative to its equilibrium position, $u_T$, reaches a
certain fraction of the spacing between the flux lines, $a$:
\begin{equation}\label{5}
 u_T^2=c_L^2a^2,
\end{equation}
where $a=(\Phi_0/H)^{1/2}$, $\Phi_0$ is the flux quantum, and
$c_L$ is the Lindemann constant. The magnitude $u_T$ depends on
the elastic moduli of the lattice \cite{25} and was calculated in
many papers; see, e.g., Refs.~\onlinecite{9},
\onlinecite{25,26,27,28}. It can be represented in the
form:\cite{4}
\begin{equation} \label{6}
  u_T^2\approx \xi^2 \cdot t
  \left ({Gi\over 1-t^2} \! \right)^{\!1/2}h^{-1/2}f(h),
\end{equation}
where $\xi(t)$ is the coherence length in the $ab$ plane,
$h=H/H_{c2}(t)$, $t=T/T_c$, $H_{c2}(t)=\Phi_0/2\pi\xi^2$ is the
upper critical field, $Gi\,$ is the Ginzburg number,
\[
  Gi={1\over 2}\left ({T_c \over H_c^2\,\epsilon\, \xi_0^3}
  \right)^{\!2},
\]
which characterizes the width of the fluctuation region in zero
magnetic field, $\xi_0$ and $H_c$ are the coherence length and
thermodynamic critical magnetic field of the superconductor in
the Ginzburg - Landau theory extrapolated to $T=0$. For
definiteness, we implied in Eq.~(\ref{6}) and below that
$\xi^2(t)=\xi^2(0)/(1-t^2)$. [Hence $\xi_0=\xi(0)/\sqrt 2$]. The
complete expression for the function $f(h)$ was given in
Ref.~\onlinecite{28}, but for our further analysis it is
sufficient to use a simplified form of this function \cite{4} in
which the contribution containing the compression modulus of
the vortex lattice, $c_{11}$, is neglected:
  \begin{equation} \label{7}
  f(h)= {2\beta_A \over 1-h} {[1+(1+\tilde c)^2]^{1/2}-1
  \over \tilde c(1+\tilde c)},
  \end{equation}
with $\tilde c=0.5[\beta_A(1-h)]^{1/2}$, and $\beta_A=1.16$. Note
that this formula can be rewritten in the form:
  \begin{equation} \label{8}
  f(h)= {f_1(h) \over (1-h)^{3/2}},
  \end{equation}
where the function $f_1(h)$ defined by this equality decreases
monotonically with increasing $h$, and its variation in the
interval $0<h<1$ is not large: $f_1(0)\approx 2.34$,
$f_1(1)\approx 1.78$. Thus, to a first approximation, this
function can be considered as a constant, $f_1(h)\approx f_1 \sim
2$.

Combining  formulas (\ref{5})-(\ref{8}), we arrive at the equation
for the normalized melting field $h_m(t)=H_m(t)/H_{c2}(t)$:
\begin{equation} \label{9}
  t \left ({Gi\over 1-t^2} \right)^{\!1/2} \!h_m^{1/2}
  \,{f_1(h_m)\over (1-h_m)^{3/2}}=2\pi c_L^2.
\end{equation}
This equation agrees with those obtained
earlier\cite{9,25,26,27,28} and differs from them only in the form
of the function $f_1(h)$ since different authors used slightly
different approximations for the elastic moduli or took into
account the contribution associated with the compression modulus
of the vortex lattice.

When the normalized melting field $h_m(t)$ is small, $h_m\ll 1$,
it follows from Eq.~(\ref{9}) that
  \begin{equation} \label{10}
  {H_m(t) \over H_{c2}(0)}=
  \left({2\pi c_L^2 \over f_1(0)t}\right)^{\!2} {(1-t^2)^2 \over
  Gi}.
  \end{equation}
This is a well-known result \cite{9,25,26,27,28} that holds for
temperatures near $T_c$ (but outside the fluctuation region in
zero magnetic field), namely for
 \begin{equation}\label{11}
 Gi \ll 1-t^2 \ll (f_1(0)/ 2\pi c_L^2)^2\, Gi.
 \end{equation}
Note that for inequalities (\ref{11}) to be fulfilled in a
sufficiently wide temperature interval, the Ginzburg number should
not be too small. In the opposite limiting case,
 \begin{equation}\label{12}
 1-t^2 \gg (f_1(1)/2\pi c_L^2)^2\, Gi,
 \end{equation}
the field $h_m$ is close to unity, and one obtains from
Eq.~(\ref{9}):
\begin{equation} \label{13}
 1-h_m\approx \left({f_1(1)\over 2\pi c_L^2}\right)^{\!2/3} \!
 t^{2/3}  \left({Gi \over 1-t^2}\right)^{\!1/3}  \!,
\end{equation}
or equivalently,
\begin{equation} \label{14}
 {H_{c2}(t)-H_m(t)\over H_{c2}(0)}\approx
 \left({f_1(1)\over 2\pi c_L^2}\right)^{\!2/3} \!\!
 t^{2/3}  Gi^{1/3} (1-t^2)^{\!2/3}  \!.
\end{equation}
Note that  $H_{c2}(0)Gi^{1/3} (1-t^2)^{\!2/3}$ is the width (along
the $H$ axis) of the fluctuation region in not too small magnetic
fields, \cite{29,30,31} $H\gg Gi\,H_{c2}(0)$, and for inequality
(\ref{12}) to hold, the Ginzburg number should not be too large.

\subsection{Energy balance}  

At melting, proliferation of dislocations occurs in the vortex
lattice. These dislocations create a network in the lattice, and
we consider a mean unit cell of this network composed of edge and
screw dislocations. The energies of these dislocations are of the
order of
\begin{eqnarray} \label{15}
 E_{\rm edge}& \sim & c_{66} a^2 l_{\parallel}, \nonumber \\
 E_{\rm screw} & \sim & (c_{44}c_{66})^{1/2} a^2 l_{\perp},
\end{eqnarray}
where $c_{66}$ and $c_{44}$ are the shear and tilt moduli of the
flux-line lattice, while $l_{\parallel}$ and $l_{\perp}$ are the
dimensions of the cell in the longitudinal and transverse
directions to $H$, respectively. Since at the melting these
dimensions are of the order of $a$ (see below), we have omitted
the logarithmic factors $\ln(l_{\parallel}/a)$, $\ln(l_{\perp}/a)$
in the above formulas for the dislocation energies. The shear and
tilt moduli may be expressed as,\cite{25,9}
\begin{eqnarray}\label{16}
 c_{66}\approx {\varepsilon_0 \over 4a^2} (1-h)^2, \nonumber \\
 c_{44}\sim {\epsilon^2\varepsilon_0 \over a^2}(1-h),
\end{eqnarray}
where $h=H/H_{c2}(t)$, $\varepsilon_0=(\Phi_0/4\pi
\lambda_{ab})^2$, and we have omitted the logarithmic factor of
type $\ln(a/\xi)$ in the tilt modulus (this nonlocal modulus
should be estimated at wave vectors $k$ of the order of
$l_{\parallel}^{-1}$, $l_{\perp}^{-1}$). The factors containing
$(1-h)$ take into account the softening of the vortex lattice near
the $H_{c2}(t)$ line.\cite{32} Minimization of the elastic energy
of the cell at its fixed volume leads to $E_{\rm edge}\sim E_{\rm
screw}$ and hence gives the relation between $l_{\parallel}$ and
$l_{\perp}$:
\begin{equation}\label{17}
 l_{\parallel}\sim l_{\perp}\left ( {c_{44} \over c_{66}}\right
 )^{1/2}.
\end{equation}
Thus, up to a numerical factor, the elastic energy of the
dislocation cell, $E_{\rm el}$, equals $E_{\rm screw}$.

At melting, the cost in the elastic energy due to the
proliferation of the dislocation network is balanced by the
entropy gain in the free energy of the flux-line lattice. We now
give a simple estimate of this gain: The vortex-lattice degrees of
freedom associated with shear undergo a change when the melting
occurs. This change occurs for lattice modes with wave-lengths
greater than $l_{\parallel}$, $l_{\perp}$. There are
$l_{\parallel}^{-1}l_{\perp}^{-2}$ modes of this type in the unit
volume of the lattice, and each of them contributes about $T$ to
the entropy gain. Thus, up to a numerical factor, the gain per
cell of the dislocation network is $T\cdot l_{\parallel}^{-1}
l_{\perp}^{-2}\cdot l_{\parallel}l_{\perp}^{2}\sim T$.

We now can write down the change of the free energy per cell at
melting:
\begin{equation}\label{18}
\Delta F\propto C(c_{44}c_{66})^{1/2} a^2 l_{\perp} - T ,
\end{equation}
where some constant $C$ is the ratio of the unknown numerical
factors in $E_{\rm el}$ and in the entropy gain. Minimization of
Eq.~(\ref{18}) with respect to the parameter $l_{\perp}$ leads to
the conclusion that this parameter should have the minimum
possible value. It is clear that this value is of the order of $a$
in the lattice. Then, taking into account that $\Delta F=0$ at the
melting, we arrive at
\begin{equation}\label{19}
 C(c_{44}c_{66})^{1/2} a^3  - T = 0 ,
\end{equation}
where this constant $C$ may be slightly renormalized as compared
to the $C$ of Eq.~(\ref{18}). Inserting Eqs.~(\ref{16}) for the
elastic moduli and assuming $\lambda_{ab}(t)/\xi(t)=\ $const, one
finds the equation for the melting line $h_m(t)$:
\begin{equation} \label{20}
  t \left ({Gi\over 1-t^2} \right)^{\!1/2}\!\!\! {h_m^{1/2}\over
  (1-h_m)^{3/2}}={\sqrt\pi C \over 4}\,.
\end{equation}

As was noted earlier (see, e.g., Refs.~\onlinecite{3},
\onlinecite{9}), formula (\ref{10}) for the melting line near
$T_c$ can be obtained from the energy balance. Here we have taken
into proper account the softening of the elastic moduli near
$H_{c2}(T)$, and now equation (\ref{20}) shows that not only
formula (\ref{10}) but also expression (\ref{13}) can be derived
by this method. Moreover, if one uses the approximation in which
the function $f_1(h)$ is a constant,\cite{33} $f_1(h)\approx f_1$,
and put $C=8\sqrt\pi c_L^2/f_1$, Eqs.~(\ref{9}) and (\ref{20})
completely coincide in the whole temperature interval.

\subsection{Some exact results} 

Within the mean-field theory, when one neglects fluctuations of
the superconducting order parameter, the melting line $H_m(T)$
coincides with the $H_{c2}(t)$ line. It is the fluctuations that
shift $H_m(t)$ downwards in the $H$-$T$ plane. As was mentioned
above, see Eq.~(\ref{14}), the Lindemann criterion shows that at
sufficiently strong magnetic fields, the distance between the
melting line and the mean-field $H_{c2}(t)$ line is comparable
with the width of the fluctuation region. But then the question
arises about the applicability of this criterion (and of the
energy balance) for determining $H_m$ in this region of the
magnetic fields since expressions (\ref{16}) for the elastic
moduli were derived in the framework of the mean-field theory
without accounting for the fluctuations. In this context, it
should be noted that one cannot confine oneself to taking into
account only the first fluctuation correction to the elastic
moduli (in the fluctuation amplitude) since inside the fluctuation
region the amplitude is large, and corrections of all orders are
essential. In particular, the renormalized elastic moduli will
vanish on a line which differs from the mean-field $H_{c2}(t)$.
However, simple considerations \cite{30,34} show that at $1-h_m
\ll 1$, strong fluctuations can only renormalize the numerical
factor in Eq.~(\ref{13}), but the dependences of $H_m$ on $t$ and
on $Gi$ remain unchanged. We now briefly outline these
considerations.

In dimensionless units the Ginzburg-Landau Hamiltonian depends on
the three parameters: $t$, $H/H_{c2}(0)$ and the Ginzburg number
$Gi$. As well known, the quadratic part (in the order parameter)
of this Hamiltonian looks like the Hamiltonian of a particle with
double electron charge in a magnetic field, and the energy
spectrum of this particle is the so-called Landau levels. It is
essential that in fields $H\gg Gi\,H_{c2}(0)$, the distance
between these levels exceeds the width of the fluctuation region.
Thus, if one expands the order parameter into the eigenfunctions
of the particle, only the modes of the order parameter
corresponding to the lowest Landau level strongly fluctuate near
the melting line, and one may retain only these modes in the
Hamiltonian to calculate the fluctuation part of the free energy
of a superconductor. It turns out that the Hamiltonian thus
obtained depends on a single combination of the parameters. In our
notations, this combination, $Q$, can be represented in the form:
\begin{equation}\label{21}
 Q={(1-h)(1-t^2)^{1/3} \over t^{2/3} Gi^{1/3} h^{2/3} },
\end{equation}
where $h=H/H_{c2}(t)$. Hence, in this region of the magnetic
fields, the free energies of the vortex liquid, $F_{\rm liq}$, and
the vortex lattice, $F_{\rm lat}$, are also determined only by
this combination, and for the melting line $h_m(t)$ we arrive at
the equation:
\begin{equation}\label{22}
 F_{\rm liq}(Q)= F_{\rm lat}(Q).
\end{equation}
The solution of this equation has the form:
\begin{equation}\label{23}
 Q={(1-h_m)(1-t^2)^{1/3} \over t^{2/3} Gi^{1/3} h_m^{2/3} }= C_1 ,
\end{equation}
where $C_1$ is some constant. Taking into account that $h_m\approx
1$, we find from equation (\ref{23}):
\begin{equation}\label{24}
 1-h_m \approx
 C_1 t^{2/3}\left ({Gi \over 1-t^2}\right )^{1/3}.
\end{equation}
It is seen that up to a numerical factor, this expression indeed
coincides with Eq.~(\ref{13}).

The above consideration  does not use any perturbation theory and
therefore is exact, but it does not yield the constant $C_1$.
Using some variants of perturbation theory (see, e.g.,
Ref.~\onlinecite{35}), approximate expressions for $F_{\rm liq}$
and $F_{\rm lat}$ were obtained in Refs.~\onlinecite{36},
\onlinecite{37}. On this basis, an equation for the melting line,
which agrees with Eq.~(\ref{23}), was derived in these papers
together with the appropriate constant $C_1$. Hikami et
al.\cite{36} estimated $C_1\approx 7$, while Li and Rosenstein
\cite{37} who used a refined expression for $F_{\rm lat}$ found
$C_1\approx 9.5$.

A consideration similar to that presented above was also applied
to the Hamiltonian of the vortex system in the London
approximation, see Ref.~\onlinecite{38}. In the region of the
magnetic fields considered here, $H\gg H_{c1}$, one has
$\lambda_{ab} \gg a$, and the Hamiltonian, as well as the free
energy of the vortex system, are determined by a single
combination of the physical parameters:
 $\epsilon \varepsilon_0 a / T$. Thus, the melting line
$H_m(T)$ is found from the equation:
\begin{equation}\label{25}
 {\epsilon \varepsilon_0(T) a(H_m) \over T}=C_2 ,
\end{equation}
with some constant $C_2$. In other words, the London approximation
leads to $H_m(t)\propto [t\lambda_{ab}^2(t)]^{-2}$. Putting
$\lambda_{ab}(t)= \lambda_{ab}(0)(1-t^2)^{-1/2}$ and taking into
account the definition of $Gi$, we arrive at the formula:
  \begin{equation} \label{26}
  {H_m(t) \over H_{c2}(0)}=
  {\pi \over (2C_2 t)^2} {(1-t^2)^2 \over  Gi}\,,
  \end{equation}
which agrees with Eq.~(\ref{10}).

To summarize, we have shown in this section that up to numerical
factors, the three different approaches lead to the {\it same
dependences} of the vortex-lattice melting field $H_m$ on the
temperature $T$ and on the Ginzburg number $Gi$, Eqs.~(\ref{10})
and (\ref{14}). Thus, after obtaining these dependences, e.g.,
from the energy balance, for the vortex lattice with pinning, one
can guess the true form of the Lindemann criterion for this
lattice. We shall use this procedure in the next section.

\section{melting of the vortex lattice with quenched disorder} 

In this section we analyze the melting line of the vortex lattice
with pinning assuming that the line is in the bundle pinning
region (where the transverse collective pinning length $R_c$ is
greater than $a$). As it follows from the figures of Sec.~IV B,
the melting line, as a rule, does {\it entirely} lie in this
region, and only in the case of sufficiently strong $\delta T_c$
pinning can it enter the single vortex pinning region.

\subsection{Energy balance} 

We now consider the influence of pinning by point defects on
the energy balance. The adjustment of the vortex system to the
pinning potential decreases the total energy of the system, and
the amount of this decrease is just of the order of the pinning
energy. It will be shown in this section that at melting, the
pinning energy in the flux-line {\it liquid}, $E_{\rm pin}^{\rm
liq}$, is noticeably {\it greater} than the pinning energy in the
flux-line {\it lattice} (Bragg glass), $E_{\rm pin}^{\rm lat}$.
Thus, there is a gain in the free energy of the vortex system,
$E_{\rm pin}$, associated with the pinning energy: $E_{\rm
pin}=E_{\rm pin}^{\rm liq}-E_{\rm pin}^{\rm lat}\sim E_{\rm
pin}^{\rm liq}$. This gain adds to the entropy gain, and equation
(\ref{18}) is modified as follows:
\begin{equation}\label{27}
\Delta F\propto C(c_{44}c_{66})^{1/2} a^2 l_{\perp}-T\Delta S
-E_{\rm pin},
\end{equation}
where the factor $\Delta S$ takes into account the effect of
pinning on the entropy gain per dislocation cell.

In order to estimate $E_{\rm pin}^{\rm lat}$, $E_{\rm pin}^{\rm
liq}$ and $\Delta S$, it is necessary in general to take into
account the so-called thermal depinning\cite{9,39} since the
thermal displacement $u_T$ is sufficiently large at the melting,
$u_T\sim c_La$. However, to explain the main ideas, we first carry
out the estimates neglecting this depinning and then generalize
the obtained results by taking it into account.

The pinning energy of the flux-line {\it lattice without
dislocations}, i.e., of the Bragg glass, in the volume equal to
the cell of the dislocation network, $E_{\rm pin}^{\rm lat}$, can
be estimated using the results of collective pinning
theory:\cite{9}
\begin{equation}\label{28}
 E_{\rm pin}^{\rm lat}\sim
 (W l_{\parallel}l_{\perp}^2 )^{1/2} u ,
\end{equation}
where $W=f_{\rm pin}^2 n \xi^2/a^2$, $f_{\rm pin}$ is the mean
elementary pinning force exerted by one point defect, $n$ is the
concentration of the defects, and the expression $(W
l_{\parallel}l_{\perp}^2 )^{1/2}$ is the mean pinning force per
cell. The displacement $u\equiv u(l_{\perp},0)\sim u(0,
l_{\parallel})$ is the rms relative shift of two line elements in
the vortex lattice separated by a distance $l_{\perp}$ transverse
to the magnetic field, or by $l_{\parallel}$ along the field. This
shift is caused by the random point defects. The magnitude of $u$
can be expressed in terms of the transverse collective pinning
length, $R_c$, at which the relative displacements of points in
the lattice are of the order of $\xi$. In particular, if the small
bundle pinning regime occurs, i.e., $R_c<\lambda_{ab}$, one
has:\cite{9}
\begin{equation}\label{29}
 u^2 \approx {\xi^2 \over 1+\ln(R_c^2/a^2)},
\end{equation}
where we have used that $l_{\perp}\sim a$. Note that according to
Eq.~(\ref{17}), the longitudinal dimension $l_{\parallel}$ is of
the order of $\epsilon a/(1-h)^{1/2}$.

At melting, proliferation of dislocations occurs, and in the
liquid vortices can adjust themselves to the pinning potential not
only via their elastic deformations as in the Bragg glass but also
via the plastic vortex-lattice displacements generated by
dislocations. This additional adjustment mechanism increases the
pinning energy in the liquid. To estimate the pinning energy per
dislocation cell in the liquid, $E_{\rm pin}^{\rm liq}$, it is
necessary to take into account that the displacements generated by
the dislocation network are essentially larger than the
displacements existing in the Bragg glass at the same temperature
and magnetic field within the scales $l_{\perp}$, $l_{\parallel}$.
Indeed, in the lattice without dislocations, one has
$u(l_{\perp},0)\sim u(0, l_{\parallel})=u<\xi$ [see
Eq.~(\ref{29})], while in the liquid the dislocations lead to
displacements of about $a>\xi$ at the boundary of the dislocation
cell. It is also essential that the displacements caused by the
dislocations within a scale $l< l_{\perp}$, $l_{\parallel}$ have a
different dependence on $l$ than the elastic displacements in the
Bragg glass. Then, we obtain the estimate:
\begin{equation}\label{30}
 E_{\rm pin}^{\rm liq}\sim
 (W l_{\parallel}l_{\perp}^2 )^{1/2} \xi .
\end{equation}
Note that although the displacements generated by the dislocations
are large and exceed $\xi$, we multiply the mean pinning force per
cell only by $\xi$ in formula (\ref{30}) since $\xi$ is the
effective range of the elementary pinning force $f_{\rm pin}$. It
follows from Eqs.~(\ref{28})-(\ref{30}) that $E_{\rm pin}^{\rm
liq}/E_{\rm pin}^{\rm lat}\sim (\xi/u) \sim
[1+\ln(R_c^2/a^2)]^{1/2}$, i.e., $E_{\rm pin}^{\rm liq}$
noticeably exceeds $E_{\rm pin}^{\rm lat}$ in the bundle pinning
regime, and hence formula (\ref{30}) gives the estimate of $E_{\rm
pin}$ in Eq.~(\ref{27}).

It is convenient to rewrite expression (\ref{30}) in terms of
$L_c$, the single-vortex collective pinning length, using the
relation, \cite{9} $f_{\rm pin}^2 n \approx \epsilon^4
\varepsilon_0^2/L_c^3$. Then one arrives at the formula for
$E_{\rm pin}$:
\begin{equation}\label{31}
 E_{\rm pin}\sim \epsilon \varepsilon_0 a\,
 [Dg_0(t)]^{3/2}h^{1/4}(1-h)^{3/4},
\end{equation}
where we have inserted the estimates for $l_{\perp}$,
$l_{\parallel}$, have taken into account that the quantity
$\varepsilon_0 \propto \lambda_{ab}^{-2}$ should lead to an
additional factor $1-h$ when $h=H/H_{c2}(t)$ tends to
unity,\cite{25} and have used the notation:\cite{4}
 \begin{equation}\label{32}
 {\epsilon \xi(t)\over L_c(t)}\equiv Dg_0(t)
 \end{equation}
with $\epsilon \xi(0)/ L_c(0)= D$. The function $g_0(t)$ is
given\cite{4} by
 \begin{equation} \label{33}
 g_0(t)=(1-t^2)^{1/2}
 \end{equation}
for $\delta l$ pinning, and by
 \begin{equation} \label{34}
 g_0(t)=(1-t^2)^{-1/6}
 \end{equation}
for $\delta T_c$ pinning. The parameter $D$ is a measure of the
pinning strength and is estimated as \cite{9}
$D \approx (j_c/j_0)^{1/2}$ where $j_c$ is the critical current
density in the single vortex pinning regime and $j_0$ is the
depairing current density, both taken at $T=0$.

Let us now estimate the entropy term in the energy balance
(\ref{27}). In Sec.~II B we obtained the expressions for the
melting line of the ideal vortex lattice, assuming that the main
contribution to the entropy gain at melting is due to
disappearance of the shear phonon modes in the liquid. The mechanism
of this disappearance is the following: A shear stress in the
liquid generates the so-called Peach-K\"{o}hler forces
\cite{40,41} exerted on dislocations; the dislocations begin to
move, and their shifts relax the shear stress in the liquid. In
the vortex system with pinning a dislocation cannot move if the
pinning force per dislocation cell, $(W l_{\parallel}
l_{\perp}^2 )^{1/2}$, exceeds the Peach-K\"{o}hler force $a
c_{66}u_{xy}l_{\parallel}$ exerted on a dislocation segment of
length $l_{\parallel}$ where $u_{xy}$ is the shear deformation
of the vortex system in the plane normal to the magnetic field.
Thus, for relaxation of the shear stress to occur, the
deformation $u_{xy}$ must exceed the critical value
\begin{equation}\label{35}
 u_{xy}^{cr}={(W l_{\parallel} l_{\perp}^2 )^{1/2}\over
 a c_{66}l_{\parallel}}\sim [Dg_0(t)]^{3/2}h^{-1/4}(1-h)^{-3/4}.
\end{equation}
Here we again have inserted the relations used in deriving
Eq.~(\ref{31}). In the volume with dimensions $L_{\perp}$ and
$L_{\parallel}$, thermal fluctuations generate shear deformations
$u_{xy}$ that can be estimated from
\begin{equation}\label{36}
 c_{66}u_{xy}^2L_{\parallel}L_{\perp}^2 \sim T .
\end{equation}
Hence, for large $L_{\perp}$ and $L_{\parallel}$, when
\[
 L_{\parallel} L_{\perp}^2 > [L_{\parallel}L_{\perp}^2]^{cr} \sim
 {T\over c_{66}(u_{xy}^{cr})^2} ,
\]
the deformations are less than the critical value; the relaxation
of the shear stress in the liquid does not occur, and these modes
do not contribute to the entropy gain (they do not differ
essentially from the appropriate modes in the Bragg glass). Then,
applying the considerations of Sec.~II B, we obtain the factor
$\Delta S$ in Eq.~(\ref{27}): $\Delta S=1-P(t,h)$, where
 \begin{equation}\label{37}
P(t,h)={l_{\parallel} l_{\perp}^2 \over
[L_{\parallel}L_{\perp}^2]^{cr}}\sim {[Dg_0(t)]^3(1-t^2)^{1/2}
\over t Gi^{1/2} h} .
 \end{equation}

As is well known,\cite{9,39} thermal fluctuations of the flux-line
lattice lead to a smoothing of the pinning potential and thereby
affect the pinning. We take into account the effect of this
thermal depinning on $E_{\rm pin}$ and $P(t,h)$ using the recipe
of Ref.~\onlinecite{9} (see page 1214 in that paper). Then an
additional factor $\xi/r_p$ appears in Eq.~(\ref{30}), while the
right hand side of Eq.~(\ref{35}) is multiplied by $(\xi/r_p)^2$
where $r_p=(\xi^2+u_T^2)^{1/2}$ is the new effective range of the
pinning force when thermal fluctuations are allowed for. This
modifies formulas (\ref{31}) and (\ref{37}) as follows:
\begin{eqnarray}\label{38}
 E_{\rm pin}\sim \epsilon \varepsilon_0 a\,
 {[Dg_0(t)]^{3/2}h^{1/4}(1-h)^{3/4}\over \Big [1+ t
  \left ( \displaystyle{{Gi\over 1-t^2}}\! \right)^{\!1/2}
  \displaystyle{{f(h) \over h^{1/2}}}\Big ]^{1/2}}\,, \nonumber \\
 P(t,h)\sim {[Dg_0(t)]^3(1-t^2)^{1/2}
 \over t Gi^{1/2} h \Big [1+ t
  \left ( \displaystyle{{Gi\over 1-t^2}}\! \right)^{\!1/2}
  \displaystyle{{f(h) \over h^{1/2}}}\Big ]^2}\,,
\end{eqnarray}
where we have used Eq.~(\ref{6}) for $u_T^2$.

Inserting formulas (\ref{38}) in Eq.~(\ref{27}), expressing $C$ as
$8\sqrt\pi c_L^2/f_1$ with a {\it constant} $f_1$, and taking into
account that $\Delta F=0$ at melting, we eventually find the
equation for the melting line $h_m(t)$:
\begin{eqnarray} \label{39}
  {F_T(t)\, h_m^{1/2}
  \over (1-h_m)^{3/2}}\,[1-P(t,h_m)]&+&  \\
   {A\,[Dg_0(t)]^{3/2}\,h_m^{1/4}\over
  \Big [(1-h_m)^{3/2}+
  F_T(t) h_m^{-1/2}\Big ]^{1/2}}
  &=&2\pi c_L^2\,, \nonumber
\end{eqnarray}
where
\begin{eqnarray}\label{40}
 P(t,h)= {B\,[Dg_0(t)]^3(1-t^2)^{1/2}
 \over t Gi^{1/2} \Big [h^{1/2}+ F_T(t)
  (1-h)^{-3/2}\Big ]^2}\,,
\end{eqnarray}
$A$, $B$ are some numerical factors, and we have used the notation
\begin{equation}\label{41}
 F_T(t)=f_1t \left ({Gi\over 1-t^2} \right)^{\!1/2},
\end{equation}
putting $f_1=2$ below. Equation (\ref{39}) generalizes
Eq.~(\ref{20}) obtained in the case of the ideal lattice. Note
that the factor $1-P=\Delta S$ in Eq.~(\ref{39}) naturally
explains the existence of the upper critical point on the melting
line. The temperature $t_{\rm up}$ of this point can be obtained
from the condition $\Delta S = 0$, or
 \begin{equation}\label{42}
 P(t_{\rm up}, h_m(t_{\rm up}))=1 \,,
 \end{equation}
i.e., the entropy jump in the vortex system at melting
vanishes at this point.

Equation (\ref{39}) is in agreement with the general conclusion
made in Ref.~\onlinecite{4}. Namely, it was stated\cite{4} that
the phase diagrams of various three dimensional superconductors
with point defects are determined only by the Ginzburg number
$Gi$, the parameter $D$ characterizing the strength of pinning,
and the function $g_0(t)$ which is defined by the type of pinning.
Other physical quantities [like $T_c$, $H_{c2}(T)$, $\epsilon$,
etc.] either determine only scaling factors or do not appear
explicitly in the appropriate equations at all. In particular, the
anisotropy $\epsilon$ is absorbed by the definitions of $Gi$ and
$D$ and does not enter in Eq.~(\ref{39}).

\subsection{Analysis of the equation} 

\begin{figure}
\includegraphics[scale=.5]{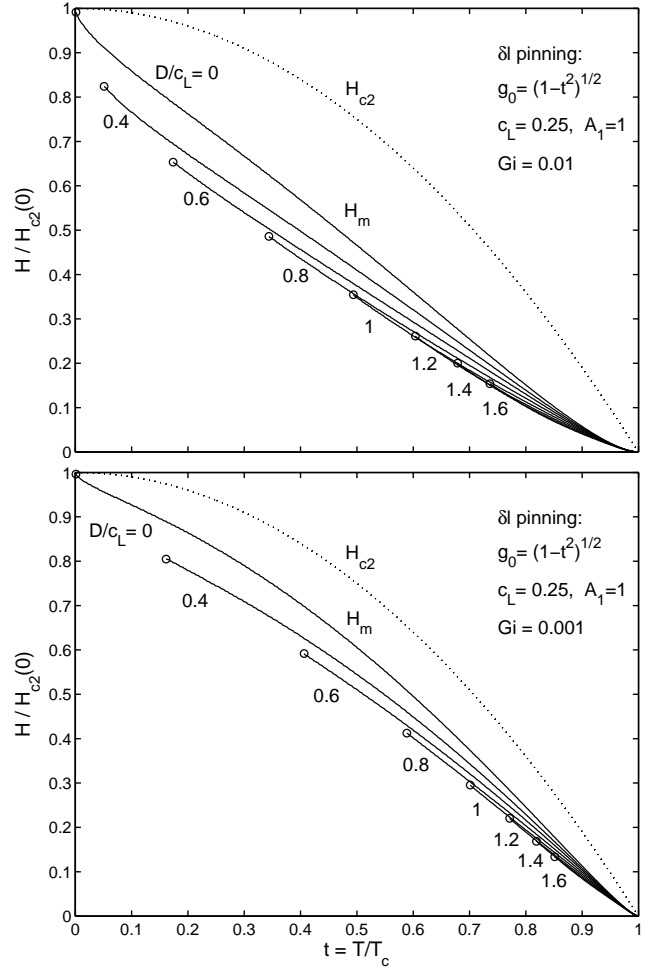}
\caption{\label{fig2} The melting line $H_m(t)$ for the vortex
lattice with quenched disorder, from Eq.~(\ref{39}), for
$D/c_L=0$, $0.4$, $0.6$, $0.8$, $1$, $1.2$, $1.4$, $1.6$ in the
case of $\delta l$ pinning at $Gi=0.01$ (top) and at $Gi=0.001$
(bottom). Here $A_1=A/(2\pi)^{3/4}=1$. The upper line $D=0$ is the
melting line of the ideal lattice, $H_m^{\rm id}(t)$. The dotted
line shows $H_{c2}(t)/H_{c2}(0) = 1-t^2$.
 }
\end{figure}

\begin{figure}
\includegraphics[scale=.5]{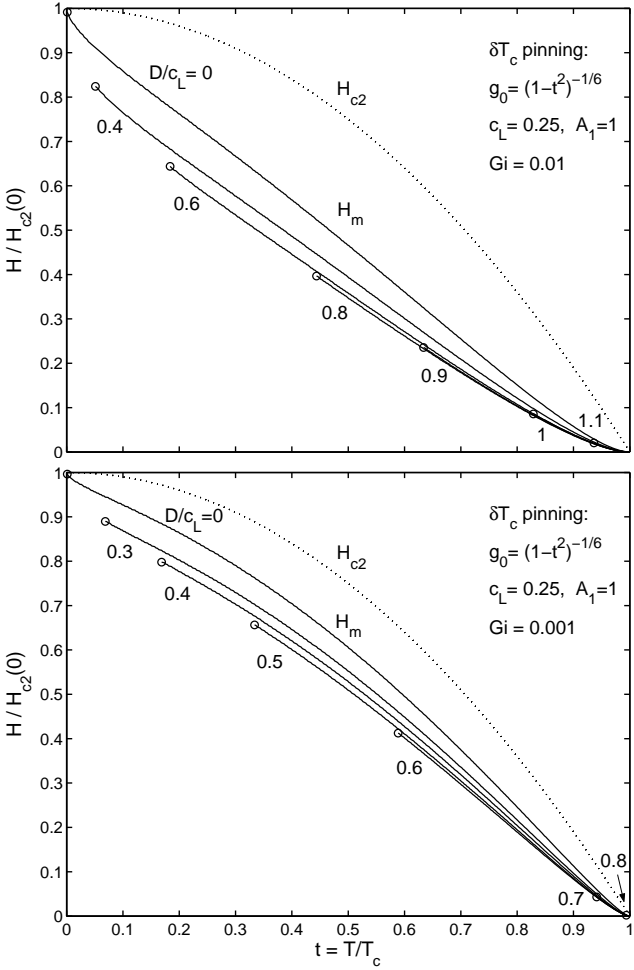}
\caption{\label{fig3} As Fig.~\ref{2} but for $\delta T_c$
pinning.
 }
\end{figure}

Figures~\ref{fig2} and \ref{fig3} present the melting lines
$H_m(t)$ obtained numerically from Eq.~(\ref{39}) for different
pinning strength $D$. Although the numerical factors $A$ and $B$
have remained unknown in the above derivation of Eq.~(\ref{39}),
the figures show that it is possible to choose
$A_1=A/(2\pi)^{3/4}$ and $B_1=Bf_1/(2\pi)^{3/2}$ so that the
calculated melting lines have the properties observed in
experiments (see Introduction). In particular, we find that the
upper critical point noticeably shifts under the influence of
pinning  and tends to $T_c$ at reasonable values of $D$ when $A_1$
and $B_1$ are not too small ($A_1$, $\sqrt B_1 \sim 1$). With
increasing $D$ the melting line shifts downwards if $B_1<A_1^2$,
and this shift becomes small when $B_1$ does not differ too much
from $A_1^2$. Of course, the choice of $A_1$ and $B_1$ on the
basis of these requirements is not unique. Let us now analyze
Eq.~(\ref{39}) in some limiting cases.

According to Sec.~II, in the temperature region (\ref{12}) one has
$1-h_m \ll 1$ for the melting line of the ideal vortex lattice. If
the parameter $D$ is not too large, this property of the melting
line remains true for the vortex lattice with pinning. But under
the condition $1-h_m\ll 1$ the lowest Landau level approximation
is valid, and we can use the results of Appendix A where the
effect of quenched disorder on the melting line has been analyzed
in the framework of an approach similar to that of Sec.~II C. On
the other hand, putting $h_m\approx 1$ in Eq.~(\ref{39}), we
arrive at
\begin{eqnarray} \label{43}
  Q^{-3/2}\!\!\left (1-{B\,f_1\,[G(t)]^2
 \over  [1+ Q^{-3/2} ]^2}\right )&+& \nonumber \\
 {A\,G(t) Q^{-3/4}\over [1+Q^{-3/2}]^{1/2}}& =&2\pi c_L^2\,,
\end{eqnarray}
where $Q^{3/2}=(1-h_m)^{3/2}/F_T(t)$ [compare with formula
(\ref{21})], and
\begin{eqnarray} \label{44}
  G(t)={[Dg_0(t)]^{3/2}\over [F_T(t)]^{1/2}}=
  {[Dg_0(t)]^{3/2}(1-t^2)^{1/4} \over (f_1t)^{1/2} Gi^{1/4}}.
\end{eqnarray}
Equation (\ref{43}) shows that the quantity $Q$ on the melting
line depends on $t$ only via the function $G(t)$. This conclusion
completely agrees with the exact result of Appendix A.

When $D=0$ [i.e., $G(t)=0$], Eq.~(\ref{43}) gives the result for
the ideal vortex lattice, $Q^{-3/2}=2\pi c_L^2$, see
Eq.~(\ref{13}). For nonzero $D$ the function $G(t)$ decreases with
the temperature. In the case of $\delta T_c$ pinning one has
$G(t)\propto t^{-1/2}$, while $G(t)$ is proportional to $(1-t^2)/
\sqrt t$ for $\delta l$ pinning. Hence in the equation for the
melting line the role of terms associated with pinning
{\it increases} with decreasing $t$. It follows from
Eqs.~(\ref{43}), (\ref{42}) that at the upper critical point
the function $G(t)$ reaches a certain value, and
condition (\ref{42}) can be written in the form:
\begin{eqnarray} \label{45}
  \sqrt {Bf_1}\, G(t_{\rm up})= {1\over 2}+\left ({1\over 4}+
  (2\pi c_L^2)^2{Bf_1 \over A^2}\right )^{1/2}.
\end{eqnarray}
Interestingly, up to a numerical factor (and the factor
$[1-h_{sv}(t)]^{3/2}$ which we do not take into account here
\cite{42}) this condition reduces to Eq.~(32) of
Ref.~\onlinecite{4} for the intersection point of the melting line
with the order--disorder line. In particular, it follows from
Eqs.~(\ref{44}), (\ref{45}) that $t_{\rm up}$, similarly to the
temperature of the intersection point, depends on $Gi$ and $D$
only via the combination $D^3/Gi^{1/2}$ of these parameters.

The so-called depinning line\cite{9} along which $u_T\approx \xi$,
intersects the melting line of the ideal lattice, $H_m^{\rm
id}(t)$, at $1-t^2\sim (f_1^2/2\pi c_L^2)\,Gi$, i.e., outside the
temperature region defined by Eq.~(\ref{12}). In the region
$1-t^2\ll (f_1^2/2\pi c_L^2)\,Gi$, the depinning is essential. In
this depinning region, the terms $T\Delta S$ and $E_{\rm pin}$ in
the energy balance (\ref{27}) become relatively small, and the
melting line is close to that of the ideal lattice.

\subsection{Lindemann criterion} 

We now rewrite Eq.~(\ref{39}) in the form of the Lindemann
criterion. As it follows from formulas (\ref{5}) and ({\ref{9}),
the factor $F_T(t) h_m^{1/2}/(1-h_m)^{3/2}$ in the first term of
Eq.~(\ref{39}) is simply $2\pi u_T^2/a^2$ where $u_T$ is
described\cite{43} by Eq.~(\ref{6}). To rewrite the second term in
Eq.~(\ref{39}), let us take into account the formula for the
averaged relative shift of two line elements in the vortex lattice
separated by a distance $R\le R_c$ transverse to the magnetic
field:\cite{44}
 \begin{equation} \label{46}
 {u^2(R,0)\over r_p^2}\approx \left ({\epsilon a\over L_c}{\xi^2
 \over r_p^2}\right )^3
  {1+ \ln(R^2/a^2)+ \epsilon R/\lambda_{ab}
 \over (1-h)^{3/2}} .
  \end{equation}
[If the small bundle pinning regime occurs, $R_c<\lambda_{ab}$,
the last term in this formula, $\epsilon R/\lambda_{ab}$, is small
and may be omitted.] At $R=R_c$, one has $u(R_c,0)=r_p$. Inserting
$r_p^2=\xi^2+u_T^2$ and the expression (\ref{6}) for $u_T$ in the
relationship thus obtained, and taking into account the definition
(\ref{32}), we arrive at
 \begin{eqnarray} \label{47}
 1+ \ln\left ({R_c^2\over a^2}\right )+ \epsilon {R_c\over
 \lambda_{ab}}
 &=& \left ({h(1-h)\over 2\pi D^2g_0^2(t)}\right )^{3/2}\times  \\
  \Big [1&+& \left ( {Gi\over 1-t^2}  \! \right)^{\!1/2}
  \!\!\!\!\!{f_1 t\over h^{1/2}(1-h)^{3/2}}\Big ]^{3}.
  \nonumber
  \end{eqnarray}
Simple manipulations using this formula show that the second term
in Eq.~(\ref{39}), which tells the role of the pinning energy in
the energy balance, is
 \begin{equation} \label{48}
 2\pi A_1 {r_p^2\over a^2} \left [1+
 \ln\left ({R_c^2\over a^2}\right )+ {\epsilon
 R_c\over \lambda_{ab}}\right ]^{-1/2},
  \end{equation}
where $A_1=A/(2\pi)^{3/4}$. Since the factor containing $R_c$ can
be represented as $u(a,0)/r_p$ [using Eq.~(\ref{46}) at $R=a$ and
then at $R=R_c$], we find one more form for the second term in
Eq.~(\ref{39}):
 \begin{equation} \label{49}
 2\pi A_1 {r_p u(a,0)\over a^2}.
  \end{equation}
As to the quantity $P(t,h)$ given by Eq.~(\ref{40}), it is
expressed as $B_1u^2(a,0)/u_T^2$ where $B_1=Bf_1/(2\pi)^{3/2}$.
Thus, equation (\ref{39}) is equivalent to the following
criterion:
 \begin{equation} \label{50}
 u_T^2-B_1 u^2(a,0)+ A_1 r_p u(a,0)=c_L^2 a^2,
  \end{equation}
which is true for the melting line in the bundle pinning region.
Equation (\ref{50}) can be also rewritten in the form:
 \begin{eqnarray} \label{51}
 u_T^2 - B_1 u^2(a,0)+ A_1 u^2(a,0) \left [1+
 \ln\left ({R_c^2\over a^2}\right )+ {\epsilon
 R_c\over \lambda_{ab}}\right ]^{1/2}\!\!\! \nonumber \\
 =c_L^2 a^2.\ \ \ \
  \end{eqnarray}
Note that although criteria (\ref{2}) and (\ref{3}) are
qualitatively close to Eq.~(\ref{51}), they underestimate pinning
at large $R_c$.

Equation (\ref{50}) [or (\ref{51})] is valid when $u_T^2\ge B_1
u^2(a,0)$ since condition (\ref{42}) for the upper critical point
now has the form:
 \begin{equation} \label{52}
 u_T^2=B_1 u^2(a,0).
  \end{equation}
Taking into account that $B_1\sim 1$, this condition means that
the upper critical point, as a rule, lies in the bundle pinning
region. Indeed, if the melting line enters the single vortex
pinning region before the upper critical point occurs, one finds
$u_T^2\ge B_1 r_p^2 = B_1 (\xi^2 + u_T^2)$ at the boundary of this
region. In other words, the melting line can intersect the
boundary in the depinning region where $u_T^2$ is large. But in
this region the melting line practically coincides with that of
the ideal lattice, Sec.~III B; the latter can cross the boundary
only in the case of $\delta T_c$ pinning at sufficiently large
$D$. \cite{19}

\section{Phase diagrams} 

\subsection{The first and the second scenarios} 

To gain some insight into the character of the mergence of the
melting line with the order--disorder line, let us consider more
closely the dependence of the free energy difference between the
liquid and the Bragg glass, $\Delta F$, on the size of the
dislocation cell, $l_{\perp}$. We shall consider only such
$l_{\perp}$ that lie near the minimum possible size in the lattice
$l_{\perp}^{\rm min}$ ($l_{\perp}^{\rm min}\sim a$). In evaluating
the entropy gain $T\Delta S$, the elastic energy $E_{\rm el}$ and
the pinning energy $E_{\rm pin}$, it is necessary to take into
account that the modulus $c_{44}$ depends on the wave vector ${\bf
k}$. According to Ref.~\onlinecite{25}, one has $c_{44}(k\sim
1/l_{\perp}) \approx c_{44}(1/a)\cdot (l_{\perp}/a)^2$ at
$l_{\perp}<\lambda_{ab}/\epsilon$, and hence $l_{\parallel}\propto
l_{\perp}^2$, see formula (\ref{17}). Then we can write $\Delta F$
in the form (we still neglect logarithmic factors):
\begin{equation}\label{53}
\Delta F\propto E_{\rm el}(a) \tilde l_{\perp}^2-E_{\rm
pin}(a)\tilde l_{\perp}^2+T[P(a)\tilde l_{\perp}^4 -1] ,
\end{equation}
where $\tilde l_{\perp}=l_{\perp}/l_{\perp}^{\rm min}$, $P(a)$
denotes $P(t,h_m)$ of Eq.~(\ref{39}), and $E_{\rm el}(a)$ and
$E_{\rm pin}(a)$ are the elastic and the pinning energies per
dislocation cell at $l_{\perp}=l_{\perp}^{\rm min}$, which have
been inserted into formula (\ref{27}) to derive Eq.~(\ref{39}).
Equation (\ref{53}) is valid when $P(a)\tilde l_{\perp}^4 -1<0$,
otherwise the last term in this equation has to be omitted, and
one has,
\begin{equation}\label{54}
\Delta F\propto E_{\rm el}(a) \tilde l_{\perp}^2-E_{\rm
pin}(a)\tilde l_{\perp}^2 ,
\end{equation}
for $\tilde l_{\perp}^4> 1/P(a)$.

As was mentioned above, in the equation (\ref{39}) for $H_m(t)$
[or equivalently $T_m(H)$] the relative role of terms associated
with pinning increases with decreasing $T_m$ for both types of
pinning. When $E_{\rm pin}(a)$ and $P(a)$ are sufficiently small
(i.e., for temperatures $T_m$ near $T_c$), the function $\Delta
F(\tilde l_{\perp})$ is minimum at the lowest possible value
$\tilde l_{\perp}=1$ and then increases with increasing $\tilde
l_{\perp}$, Fig.~\ref{fig4}a. This means that melting occurs at
$\tilde l_{\perp}=1$ as in the ideal lattice. This conclusion has
been already used implicitly in deriving Eq.~(\ref{39}). When the
temperature $T_m$ decreases, and hence $P(a)$ and the ratio
$E_{\rm pin}(a)/E_{\rm el}(a)$ increase, the point $\tilde
l_{\perp} = [1/P(a)]^{1/4}$ shifts to smaller $\tilde l_{\perp}$,
and the slope of $\Delta F$ in Eq.~(\ref{54}) on $\tilde
l_{\perp}^2$ decreases. Eventually, at the critical temperature
$t_{\rm up}$, one arrives at the situation shown either in
Fig.~\ref{fig4}c or in Fig.~\ref{fig4}d since the energy balance
for the melting $E_{\rm el}(a) - E_{\rm pin}(a) =T_m[1-P(a)]$
leads to $E_{\rm el}(a) - E_{\rm pin}(a)=0$ at this point. In the
case of Fig.~\ref{fig4}d when the dependence of $\Delta F$ on
$\tilde l_{\perp}^2$ has a negative curvature at $\tilde
l_{\perp}=1$, we conclude that this case is necessarily preceded
by the situation shown in Fig.~\ref{fig4}b, which has to occur at
some temperature $t_i > t_{\rm up}$ where $P(a)<1$ and $E_{\rm
pin}(a) < E_{\rm el}(a)$. At this $t_i$ the size of the
dislocation cell, $l_{\perp}$, sharply increases because the
absolute minimum of $\Delta F$ now occurs at a larger $\tilde
l_{\perp}>1$. In other words, at this temperature we find an {\it
intersection} of the melting line with the order--disorder line.
The abrupt increase of $l_{\perp}$ also means that the melting and
the order--disorder transition cannot form a unified phase
transition line. Thus, we arrive at the first scenario\cite{45}
shown in Fig.~\ref{fig1}. In the case of Fig.~\ref{fig4}c two
possibilities exist: Either at some temperature $t_i>t_{\rm up}$
the situation shown in Fig.~\ref{fig4}b occurs, and we again
arrive at the first scenario, or the function $\Delta F(\tilde
l_{\perp})- \Delta F(1)$ remains positive for all $\tilde
l_{\perp}$ in the temperature interval $t_{\rm up} \le t \le 1$.
In the latter case we find that the mean size of the dislocation
cell, $l_{\perp}$, cannot gradually change in the process of the
reduction of the entropy gain, and hence one has $\tilde
l_{\perp}=1$ down to $t_{\rm up}$. In this case, the melting line
continuously transforms into the order--disorder line, and we
arrive at the second scenario. Moreover, since $t_{\rm up}$ is not
a specific temperature for $E_{\rm pin}(a)$ or $E_{\rm el}(a)$, it
is quite probable that the result $\tilde l_{\perp}=1$ remains
true also at $t<t_{\rm up}$ (at least in the bundle pinning
region). Then, at $t\le t_{\rm up}$ the order--disorder line
(which is now a part of the unified phase transition line) is
determined by the condition,
\begin{equation}\label{55}   
E_{\rm el}(a)=E_{\rm pin}(a) .
\end{equation}

As it follows from the results of Sec.~III C, in the bundle
pinning region condition (\ref{55}) is equivalent to the Lindemann
criterion:
 \begin{eqnarray} \label{56}  
 A_1 u^2(a,0) \left [1+
 \ln\left ({R_c^2\over a^2}\right )+ {\epsilon
 R_c\over \lambda_{ab}}\right ]^{1/2}\!\!\!
 =c_L^2 a^2.\ \ \ \
  \end{eqnarray}
If the order--disorder line enters the single vortex pinning
region, where $R_c=a$, condition (\ref{55}) yields:
 \begin{eqnarray} \label{57}  
 A_1 u^2(a,0)  = c_L^2 a^2.\ \ \ \
  \end{eqnarray}
(Note that the ratio $\epsilon a/ \lambda_{ab}$ is negligible.) In
Appendix B equations (\ref{56}) and (\ref{57}) are presented in
explicit form.

 \begin{figure}
\includegraphics[scale=.5]{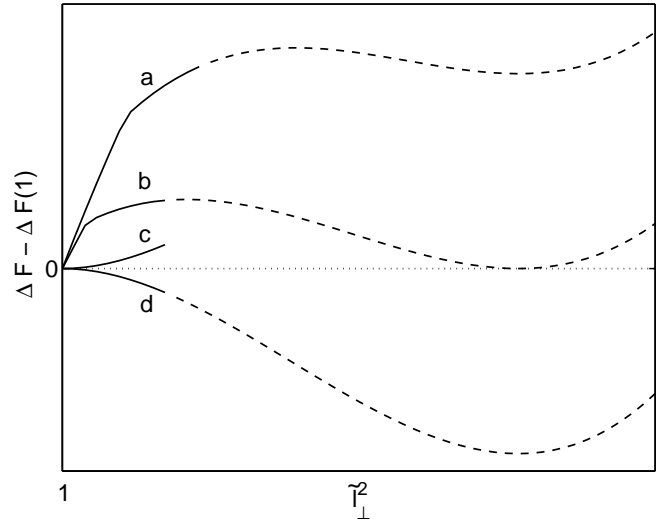}
\caption{\label{fig4} Schematic sketch of the function $\Delta
F(\tilde l_{\perp})$, Eqs.~(\ref{53}), (\ref{54}), for $t>t_i$
(a), $t=t_i$ (b), and $t=t_{\rm up}$ (c or d). The parts of the
curves where Eqs.~(\ref{53}), (\ref{54}) fail are indicated as
dashed lines. The breaks of the curves a and b occur at the points
$\tilde l_{\perp}^4=1/P(a)$ where Eq.~(\ref{53}) transforms into
Eq.~(\ref{54}).
 }
\end{figure}

To calculate the order--disorder line $H_{dis}(t)$ in the case of
the first scenario, we used the criterion:\cite{4}
\begin{equation} \label{58}
  u^2(a,0)=c_{LD}^2a^2 ,
\end{equation}
where $c_{LD}$ is some new Lindemann constant describing the
order--disorder transition. It should be emphasized that although
Eq.~(\ref{58}) is similar to Eq.~(\ref{57}), the Lindemann
constant $c_{LD}$ in Eq.~(\ref{58}) is independent of the constant
$c_L$ defining the melting, while in the case of the second
scenario Eq.~(\ref{57}) leads to a relationship between the
Lindemann constants defining the melting and the order--disorder
lines (the ratio of these constants is $\sqrt A_1$). Criterion
(\ref{58}) both in the single vortex and in the {\it bundle
pinning}\cite{46} regions were analyzed in Ref.~\onlinecite{4},
allowing for a smoothing of the pinning potential by thermal
fluctuations. The appropriate equations in the explicit form are
presented in Appendix B. For definiteness, in the analysis below
we shall choose the constant $c_{LD}$ for the order--disorder line
as equal to the constant $c_L$ for the melting line.

\subsection{Numerical results}

Although in this paper we have not analyzed the order--disorder
line in detail, criteria (\ref{56}), (\ref{57}), and (\ref{58})
enable one to evaluate the location of this line in the $T$ - $H$
plane for the first and the second scenarios. In Figs.~\ref{fig5}
- \ref{fig8} we compare the $T$-$H$ phase diagrams of type-II
superconductors with different types of pinning for these two
scenarios. In the construction of the figures, we use the values
of the Ginzburg number $Gi=0.01$, $0.0001$ which are typical for
high-$T_c$ superconductors, and we take into account the factor
containing $[1-h_{sv}(t)]$ that has been omitted in
Sec.~III.\cite{42} The complete set of the appropriate equations
is given in Appendix B.

In Figs.~\ref{fig5} - \ref{fig8} we choose the constants $A_1$ and
$B_1$ so that the following two requirements are satisfied: First,
the properties of $H_m(t)$ observed in experiments are reproduced,
namely, with increasing $D$ the upper critical point clearly
shifts, while the downward shift of the melting line is
small, see Sec.~III B. Second, the upper critical point $t_{\rm
up}$ of the melting line coincides with the intersection point
$t_i$ of this line with the order--disorder line (calculated
within the first scenario) at $D=0.7$ for the case of
$\delta l$ pinning. The latter requirement is due to the following
considerations: According to Ref.~\onlinecite{12},
\onlinecite{15}, the coincidence of $t_i$ and $t_{\rm up}$ is
observed in overdoped YBa$_2$Cu$_3$O$_y$ crystals ($y >6.92$)
for which the upper critical point lies at sufficiently large
magnetic fields, and so the case $h_m \sim 1$ appears to occur
there. As it was mentioned in Sec.~III B, in this case $t_i$ and
$t_{\rm up}$ depend on $Gi$ and $D$ only via a single
combination of these parameters, $D^3/Gi^{1/2}$. Therefore, one
may expect that if the coincidence of the upper critical point
with the intersection point occurs, it practically will not depend
on the specific choice of $D$ or $Gi$. In other words, the
coincidence will approximately occur for different $D$, $Gi$, and
types of pinning as long as $h_i\approx 1$. Although $h_i$ is not
too close to unity in Figs.~\ref{fig5} - \ref{fig7} (top), the
data of these figures support this statement. As one might expect,
the region of the coincidence is especially wide in $D$ for small
$Gi$, Fig.~\ref{fig6}. Thus, we have introduced the second
requirement here in order to fit the phase diagrams calculated in
the framework of the first scenario to the experimental situation
observed in the overdoped YBa$_2$Cu$_3$O$_y$ crystals.

The presented figures show that when the intersection point is
close to the upper critical point, both scenarios lead to
qualitatively similar phase diagrams. However, when the strength
of pinning $D$ increases, and the intersection point shifts toward
$T_c$, the coincidence of the points fails especially for $\delta
T_c$ pinning. In this case the first scenario leads to a
noticeable extension of the melting line beyond the
intersection point. Thus, if the first scenario really occurs,
this result possibly explains the experimental findings
\cite{11,12,13,14,15} for optimally doped YBa$_2$Cu$_3$O$_y$
crystals ($y \approx 6.92$) in which the upper critical point lies
at larger magnetic field than the intersection point. Note also
that in the case of $\delta T_c$ pinning, the melting line enters
the single vortex pinning region near $T_c$ at
$\nu \equiv (2\pi)^{3/2} D^3/Gi^{1/2}$
of the order of several units.\cite{4,19}

In Figs.~\ref{fig5} - \ref{fig8} we also show the boundary of the
single vortex pinning region, $H_{sv}(t)$. It should be noted that
apart from the well-known lower single vortex pinning region,
\cite{9} an upper region exists where this pinning occurs. The
upper region of single-vortex pinning was discussed by Larkin and
Ovchinnikov \cite{47} in the context of the origin of the peak
effect in low-$T_c$ superconductors. Without account of thermal
fluctuations, this region adjoins to the $H_{c2}(t)$ line and is
caused by the softening of the vortex lattice near $H_{c2}(t)$.
However, this softening also leads to an increase of $u_T$, which
reduces the strength of pinning. As a result of these two opposite
tendencies, the upper region does not extend to $T_c$ and has the
shape of a ``tongue''.\cite{4} Interestingly, when the strength of
pinning increases, the upper and lower single-vortex pinning
regions can merge at low temperatures. This merging occurs for
$D\ge (4\pi)^{-1/2}$; this value is independent of $Gi$ and the
type of pinning since the merging starts at $T=0$.

 \begin{figure}
\includegraphics[scale=.45]{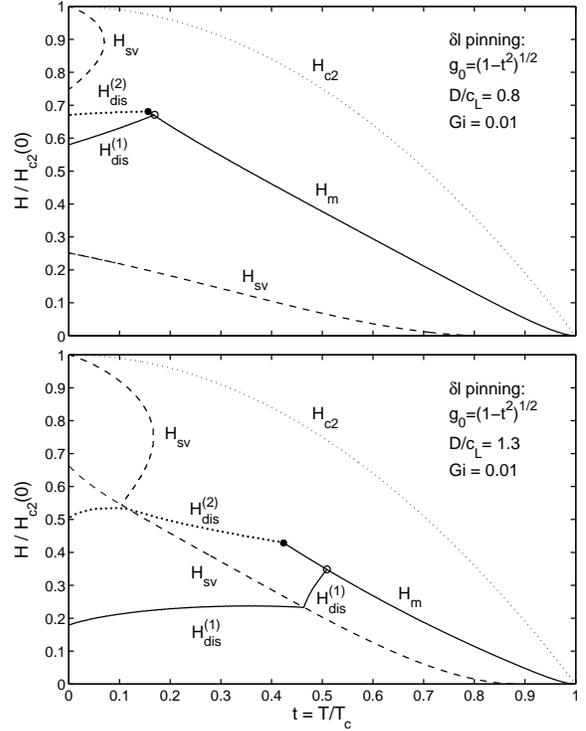}
\caption{\label{fig5}The phase diagram in the case of $\delta l$
pinning, $g_0(t)=(1-t^2)^{1/2}$. Here $A_1=0.66$, $B=0.8A_1^2$,
$c_L=0.25$, $Gi=0.01$, and $D/c_L=0.8$ (top) or $D/c_L=1.3$
(bottom). The melting line $H_m(t)$ and the order--disorder line
$H_{dis}^{(1)}(t)$ for the first scenario are shown by solid
lines, while the thick dotted line gives the order--disorder line
$H_{dis}^{(2)}(t)$ for the second scenario. The dashed line
depicts the boundary of the single vortex pinning region,
$H_{sv}(t)$, Eq.~(B2) [the thin-dashed line shows Eq.~(B1)].
The dotted line is $H_{c2}(t)/H_{c2}(0) = 1-t^2$.
The upper critical point ($T_{\rm up}$, $H_{\rm up}$) is marked by
a dot and the intersection point ($T_i$, $H_i$) by a circle.
 }
\end{figure}

 \begin{figure}
\includegraphics[scale=.45]{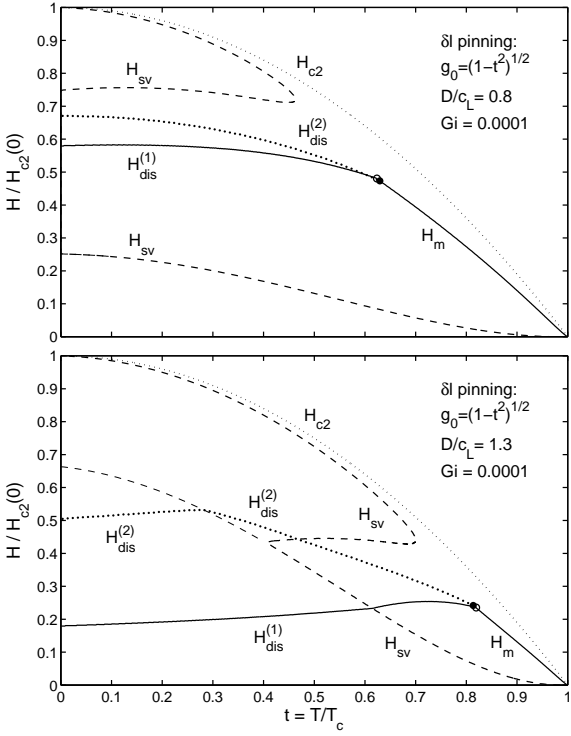}
\caption{\label{fig6} As Fig.~\ref{fig5}, but for $Gi=0.0001$.
 }
\end{figure}

 \begin{figure}
\includegraphics[scale=.45]{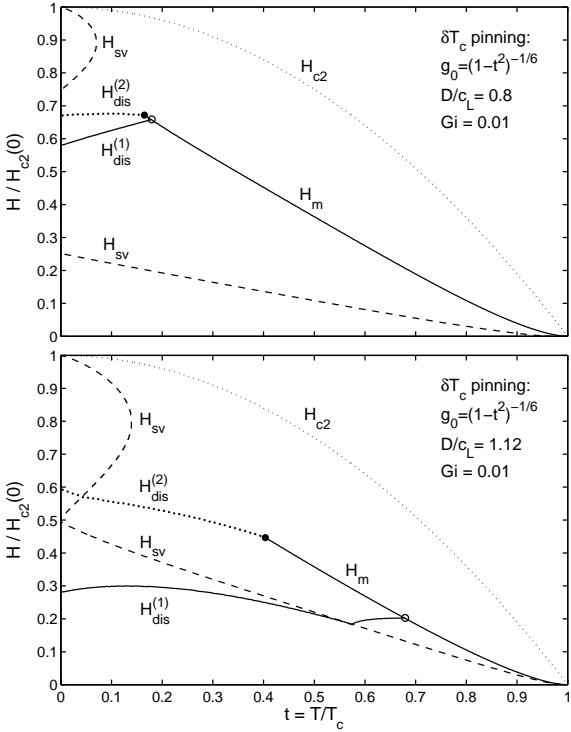}
\caption{\label{fig7} As Fig.~\ref{fig5}, but for the case of
$\delta T_c$ pinning, $g_0(t)=(1-t^2)^{-1/6}$. Top: $D/c_L=0.8$;
bottom: $D/c_L=1.12$.
 }
\end{figure}

 \begin{figure}
\includegraphics[scale=.45]{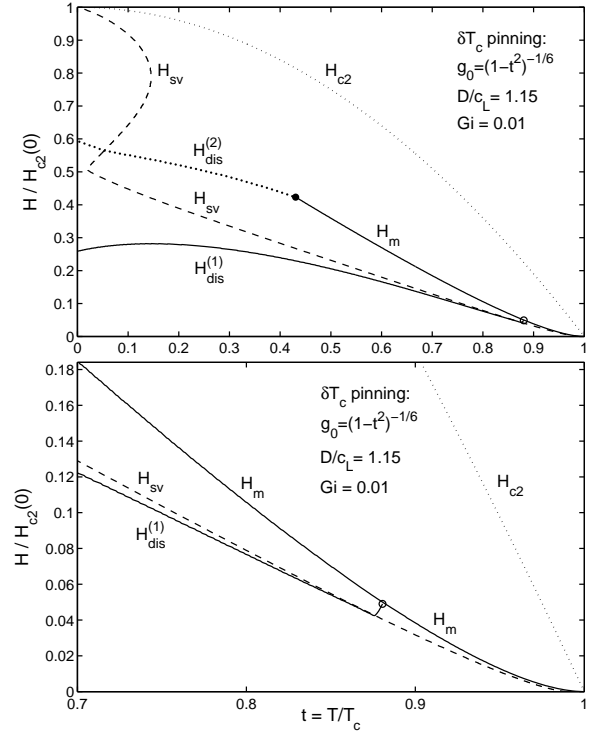}
\caption{\label{fig8} As Fig.~\ref{fig7}, but for $D/c_L=1.15$.
The lower panel shows the same phase diagram near $T_c$ at
enlarged scale.
 }
\end{figure}

The authors of Ref.~\onlinecite{5} argued that only the second
scenario can explain the decrease of $H_{dis}$ with the temperature
that was observed in their experiments. However, the presented
figures show that the order--disorder line found from criterion
(\ref{58}) (the first scenario) can decrease or increase with
temperature and even can be nonmonotonic. This depends on whether
the line is in the single vortex pinning region or in the bundle
pinning region and also on the Ginzburg number. It should be noted
that in contrast to Eq.~(\ref{3}) we take into account only the
displacement caused by the quenched disorder and do {\it not}
include
$u_T$ explicitly in the equation for the order--disorder line,
compare Eq.~(\ref{58}) with Eq.~(\ref{3}). However, the thermal
depinning (which depends on $u_T$), the softening of the elastic
moduli, and the possibility that the order--disorder line lies
not only in the single vortex pinning region but also outside it,
already produce the depicted variety of shapes of $H_{dis}(t)$.
Hence, even in the framework of the first scenario the presented
results can explain the fact \cite{5} that the order--disorder
lines observed in experiments have various shapes.

Finally, we briefly discuss the case of low-$T_c$ superconductors,
which have a very small Ginzburg number. In this case one has
$D^3/Gi^{1/2} \gg 1$ even for weak pinning strength $D$.
Since the temperatures $T_i$ and $T_{\rm up}$ are mainly
determined by this ratio of the parameters and increase when the
ratio increases, we find that in low-$T_c$ superconductors these
temperatures practically coincide with $T_c$, and only the
order--disorder line can be observed.\cite{48} Interestingly, in
this case the order--disorder lines for the first and for the
second scenarios have the same functional dependences on $t$ and
on the parameters $D$, $Gi$ [this follows from equations
(\ref{B5}) and (\ref{B8})], and moreover, they practically
coincide with each other when $A_1^2\approx 2\pi c_L^2$.

\acknowledgments

  This work was supported by the German Israeli Research Grant
Agreement (GIF) No G-705-50.14/01.

\appendix

\section{Effect of disorder on the melting line: analysis
beyond perturbation theory.}   

Assuming $1-t\ll 1$, we shall consider the partition function of
the vortex system as a functional integral with the
Ginzburg-Landau Hamiltonian.\cite{30,34} In dimensionless units
this Hamiltonian has the form:
 \begin{eqnarray}\label{A1}
 {H_{GL} \over T}= \int d{\bf r} \Big [ |\partial_z
 \psi|^2+|(-i\nabla
 +{\bf A})\psi|^2 + \tau |\psi|^2+  \nonumber \\
 2^{3/2}\pi Gi^{1/2}t|\psi|^4 \Big ],
 \end{eqnarray}
where $\psi$ is the order parameter; $\tau=t-1$; the coordinates
$x$ and $y$ are measured in units of $\xi_0$, and $z$ in units of
$\epsilon \xi_0$; $\xi_0$ is the zero-temperature coherence length
in the Ginzburg-Landau theory; the magnetic field is $b=2\pi
\xi_0^2 H/\Phi_0$ and ${\bf A}$ is its vector potential;
$\nabla=(\partial_x,\partial_y)$. Consider the melting line
$h_m(t)$ in the temperature region described by Eq.~(\ref{12}) in
which $1-h_m \ll 1$, and  hence the lowest Landau level
approximation is valid (see Sec.~II C). In this approximation the
second term of Eq.~(\ref{A1}) reduces to
 \begin{equation}\label{A2}
 b|\psi|^2 .
 \end{equation}
Pinning is introduced into the {\it dimensional} Ginzburg -Landau
Hamiltonian either via spatial disorder in the transition
temperature $T_c+\delta T_c({\bf r})$ ($\delta T_c$ pinning) or by
spatial variation of the effective mass $m+ \delta m({\bf r})$
describing disorder in the mean free path $l$ of quasiparticles
($\delta l$ pinning).\cite{9} Here $m$ is the effective mass in
the $x$-$y$ plane (since the magnetic field is along the $z$
axis). Thus, with quenched disorder in the vortex lattice, one
should add to the dimensionless Hamiltonian (\ref{A1}) the term
 \begin{equation}\label{A3}
 \varphi ({\bf r})|\psi|^2,
 \end{equation}
where
 \begin{equation}\label{A4}
 \varphi ({\bf r})= {\delta T_c({\bf r})\over T_c}
 \end{equation}
in the case of $\delta T_c$ pinning, and
 \begin{equation}\label{A5}
 \varphi ({\bf r})={\delta m({\bf r})\over m} \,b \approx
 {\delta m({\bf r}) \over m} \, |\tau|
 \end{equation}
in the case of $\delta l$ pinning. In Eq.~(\ref{A5}) we have
used the relation
 \begin{equation}\label{A6}
 b=|\tau|h
 \end{equation}
that follows from the definitions of $b$ and $h$ and put $h=1$
since $1-h\ll 1$ in the lowest Landau level approximation. For
pinning by point defects, it is assumed in the collective pinning
theory \cite{9} that disorder in $\delta T_c({\bf r})$ and in
$\delta m({\bf r})$ is short scale and described by a Gaussian
distribution with zero mean value, $ \langle \delta T_c({\bf r})
\rangle = \langle \delta m({\bf r}) \rangle = 0$, and with the
correlation function: \cite{A}
 \begin{equation}\label{A7}
 {\langle \delta T_c({\bf r}) \delta T_c({\bf r'}) \rangle
 \over T_c^2} = 2\pi D^3 \delta({\bf r}  - {\bf r'})
 \end{equation}
for $\delta T_c$ pinning, and
 \begin{equation}\label{A8}
 {\langle \delta m({\bf r}) \delta m({\bf r'}) \rangle \over m^2}
 = {30\pi \over 7} D^3 \delta({\bf r}  - {\bf r'})
 \end{equation}
for $\delta l$ pinning. Here $\langle \dots \rangle$ means disorder
averaging. Note that in agreement with Ref.~\onlinecite{4} and
with Sec.~III A of this paper, equations (\ref{A1}) - (\ref{A8})
again show that the phase diagrams of type-II superconductors with
point defects depend only on $D$ and $Gi$.

To proceed further, let us rescale the coordinates and the order
parameter $\psi$ similarly to Ref.~\onlinecite{35}:
 \begin{eqnarray}
 \tilde x = \sqrt b\, x, \label{A9} \\
 \tilde y =\sqrt b\, y, \label{A10} \\
 \tilde z = z\,Gi^{1/6} (tb)^{1/3}, \label{A11} \\
 \tilde \psi^2=\psi^2 Gi^{1/6} t^{1/3} b^{-2/3}, \label{A12}
 \end{eqnarray}
where $\tilde x$, $\tilde y$, $\tilde z$, $\tilde \psi$ are the
new coordinates and order parameter. Beside this, to agree with
the notation used in the main text of this paper, we put
$\tau=t^2-1$ below. Then, the Hamiltonian (\ref{A1}) transforms
into:
 \begin{eqnarray}\label{A13}
 {H_{GL} \over T}= \int d\tilde {\bf r} \Big [ |\partial_{\tilde z}
 \tilde \psi|^2-Q|\tilde \psi|^2+2^{3/2}\pi |\tilde \psi|^4 \Big
 ],
 \end{eqnarray}
where $Q$ is defined by Eq.~(\ref{21}). In the case of the ideal
lattice, Eq.~(\ref{23}) for $h_m$ follows from this Hamiltonian,
see Sec.~II C. In a lattice with quenched disorder, the
additional term from (\ref{A3}) in the Hamiltonian has the form:
 \begin{equation}\label{A14}
 \tilde \varphi (\tilde {\bf r})|\tilde \psi|^2,
 \end{equation}
with $ \langle\tilde \varphi(\tilde {\bf r}) \rangle =0$ and
 \begin{equation}\label{A15}
 \langle \tilde \varphi (\tilde {\bf r})  \tilde \varphi
 (\tilde {\bf r}')\rangle  ={2\pi D^3\over  Gi^{1/2} t}\,
 \delta(\tilde {\bf r}  - \tilde {\bf r}')
 \end{equation}
in the case of $\delta T_c$ pinning, and with
 \begin{equation}\label{A16}
 \langle \tilde \varphi (\tilde {\bf r})  \tilde \varphi (\tilde {\bf
 r}')\rangle  ={30\pi D^3 (1-t^2)^2\over 7 Gi^{1/2} t}\,
 \delta(\tilde {\bf r}  - \tilde {\bf r}')
 \end{equation}
in the case of $\delta l$ pinning.

Up to numerical factors, the right hand sides of Eqs.~(\ref{A15})
and (\ref{A16}) coincide with the function $[G(t)]^2$ defined by
Eq.~(\ref{44}). Then, the free energies of the vortex liquid,
$F_{\rm liq}$, and the vortex lattice, $F_{\rm lat}$, are
determined by $Q$ and $G(t)$, and along the melting line $h_m(t)$,
one has:
\begin{equation}\label{A17}
 F_{\rm liq}(Q, G(t))= F_{\rm lat}(Q, G(t)).
\end{equation}
Thus, on the melting line, the quantity $Q$ is some function of
$G(t)$. Note that the functional form of Eq.~(\ref{43}) agrees
with this conclusion, see Sec.~III B.

\section{Equations for calculation of the phase diagrams} 

In section III where we considered the melting line in the bundle
pinning region, the normalization factor containing
$[1-h_{sv}(t)]$ was omitted.\cite{42} The origin of this factor is
due to the additional power of $1-h$ in the shear modulus $c_{66}$
as compared with the tilt modulus $c_{44}$, see Eqs.~(\ref{16}).
In the single vortex pinning region where the shear modulus does
not play any role, the additional power should not manifest itself
in physical properties. Therefore, if a physical quantity in the
bundle pinning region contains a factor $(1-h)^n$ caused by this
additional power, it is necessary to introduce a normalization
factor $[1-h_{sv}(t)]^{-n}$ in this quantity to provide its
continuity at the boundary of the single vortex pinning region
$h_{sv}(t)$. (Here $n$ is some power.) Although the effect of the
factor $[1-h_{sv}(t)]^{-n}$ on the melting line $H_m(t)$ and on
the order--disorder line $H_{dis}(t)$ in the bundle pinning regime
is small, it essentially influences the boundary of the single
vortex pinning region, $H_{sv}(t)$, and the line $H_{dis}(t)$
inside this region. In this Appendix B, taking into account this
factor,\cite{B} we compile the complete set of equations used for
the construction of Figs.~\ref{fig5} - \ref{fig8}.

The boundary of the single vortex pinning region
$h_{sv}(t)=H_{sv}(t)/H_{c2}(t)$ is described by
equation (19) of Ref.~\onlinecite{4}:
  \begin{eqnarray}\label{B1}   
   h_{sv}^{1/2}(t)+ {F_T(t) \over
 [1-h_{sv}(t)]^{3/2}}=(2\pi )^{1/2}Dg_0(t)\, ,
 \end{eqnarray}
where we have used the notation (\ref{41}). The boundary of the
upper single vortex pinning region, discussed in Sec.~IV B, can be
obtained from the equation: \cite{4}
  \begin{eqnarray}\label{B2}  
 {1-h \over 1-h_{sv}(t) } \left[h^{1/2} \!+\! {F_T(t) \over
 (1-h)^{3/2}} \right]^2 \!\!
 =  2\pi  \left (D g_0(t) \right )^{2} \! ,
\end{eqnarray}
which in addition reproduces the root of Eq.~(\ref{B1}), i.e., it
yields the entire boundary of  the single vortex pinning regions.
Inside this upper region the vortex lattice is in a state where
$R_c=a$ and $u(a,0)=r_p$, i.e., a borderline state between the
single vortex pinning and bundle pinning regimes occurs there.

Equation (\ref{39}) for the melting line in the bundle pinning
region is rewritten as follows:
\begin{eqnarray} \label{B3}   \nonumber          
 {F_T(t)h_m^{1/2}\over (1-h_m)^{3/2}}\left [ 1-
  P(t,h_m) \right ] &+& \\
  {A\,[Dg_0(t)]^{3/2}\,h_m^{1/4}[1-h_{sv}(t)]^{3/4} \over
  \Big [(1-h_m)^{3/2}+ F_T(t)\,h_m^{-1/2}\Big ]^{1/2}}
  &=& 2\pi c_L^2\,,
\end{eqnarray}
with
 \begin{equation} \label{B4}  
 P(t,h_m)=  {B[Dg_0(t)]^3(1-t^2)^{1/2}[1-h_{sv}(t)] \over
  t\,Gi^{1/2}\left[h_m^{1/2}+
  \displaystyle{F_T(t)\over (1-h_m)^{3/2}}\right]^2}\,.
\end{equation}
Equation (\ref{B3}) is valid in the interval $t_{\rm up}\le t\le
1$ where the temperature $t_{\rm up}$ defines the position of the
upper critical point of the melting line and is given by the
condition (\ref{42}). Note that in this paper we consider
situations when the upper critical point lies in the bundle
pinning region.

In the framework of the {\it second scenario} the equation for the
order--disorder transition line (which is the continuation of the
melting line to $t < t_{\rm up}$) in the bundle pinning region
follows from criterion (\ref{56}) as:
\begin{eqnarray}\label{B5}  
  {A\,[Dg_0(t)]^{3/2}\,h_m^{1/4}[1-h_{sv}(t)]^{3/4} \over
  \Big [(1-h_m)^{3/2}+ F_T(t)\,h_m^{-1/2}\Big ]^{1/2}}
  = 2\pi c_L^2\,,
\end{eqnarray}
i.e., the first term in Eq.~(\ref{B3}) disappears. If this
order--disorder line enters the single vortex pinning regions, it
is described by equations that result from criterion (\ref{57}).
This criterion  reads in explicit form
\begin{eqnarray} \label{B6}   
   A_1\,h^{3/5}_{sv}(t)\,h_m^{2/5}
  \Big [1 + {F_T(t)\over h_{sv}^{1/2}(1-h_m)^{3/2}}\Big ]
  =2\pi c_L^2
\end{eqnarray}
for the order--disorder line  in the {\it lower} single vortex
pinning region, while in the {\it upper} single vortex pinning
region one finds:
\begin{eqnarray} \label{B7}  
 A_1 F_T(t){ h_m^{1/2} \over (1-h_m)^{3/2}}+
   A_1\,h_m = 2\pi c_L^2\,.
\end{eqnarray}
Here we have used that $u(a,0)=r_p$ in the upper region.

In the framework of the {\it first scenario}, equations for the
order--disorder line follow from condition (\ref{58}).\cite{4}
With the use of our approximation $f_1=2$, equation (24) of
Ref.~\onlinecite{4} for the order--disorder line $H_{dis}(t)$ in
the bundle pinning region can be written in the form:
 \begin{equation}\label{B8}  
 h_{dis}^{1/2}(1-h_{dis})^{3/2}=F_T(t)K_{\pm}(t),
 \end{equation}
where $h_{dis}=H_{dis}(t)/H_{c2}(t)$,
 \begin{equation}\label{B9}  
 K_{\pm}(t)=  G_1^2-1 \pm
  \left [(G_1^2-1)^2-1 \right ]^{1/2},
  \end{equation}
and
 \begin{equation} \label{B10} 
  G_1= {\pi^{1/4} G(t)[1-h_{sv}(t)]^{3/4}
  \over 2^{1/4} c_L }\,
 \end{equation}
with $G(t)$ from Eq.~(\ref{44}). The order--disorder line
$H_{dis}(t)$ in the single vortex pinning region is given by
Eq.~(22) of Ref.~\onlinecite{4}:
 \begin{eqnarray} \label{B11}  
 h_{dis}\! \left[1 + {F_T(t) \over
   (1-h_{dis})^{3/2}
  [h_{sv}(t)]^{1/2}}\right]^{5/2}\!\!\!\!\!= \nonumber \\
 2\pi c_L^2\!\!\left({2\pi c_L^2\over
 h_{sv}(t)}\right)^{3/2}.
  \end{eqnarray}
Note that apart from the factor $A_1$ Eq.~(\ref{B11}) is
equivalent to Eq.~(\ref{B6}).


\begin{thebibliography}{9}
\bibitem{31} The width of the fluctuation region adjoining the line
             $H_{c2}(t)$ is different for $H<Gi\,H_{c2}(0)$ and for
             $H>Gi\,H_{c2}(0)$, see Ref.~\onlinecite{30}.
\bibitem{42} In Ref.~\onlinecite{4},
             the normalization factor $(1-h_{sv})^{3/2}$ [with
             $h_{sv} = H_{sv}(t)/H_{c2}(t)$] was introduced
             to provide a smooth crossover of the correlation
             function Eq.~(\ref{46}) at the boundary of the single
             vortex pinning region $H_{sv}(T)$. In Sec.~III we
             deal only with the bundle pinning region and so omit
             this factor (this has only a small effect on our
             results). But we shall take into account this factor
             in the calculations of phase diagrams in Sec.~IV B,
             see formulas in Appendix B.
\bibitem{43} Note that the magnitude of the thermal fluctuations
             in the {\it ideal} vortex lattice, Eq.~(\ref{6}),
             practically coincides with the magnitude of these
             fluctuations in the lattice with pinning when the
             bundle pinning regime occurs.\cite{4}
             This is due to the fact that the main contribution
             to $u_T$ results from the modes with short wavelengths
             ($k_{\perp}\sim 1/a$), while the quenched disorder
             essentially distorts the lattice on the scale $R_c$.
\bibitem{46} Note that in Refs.~\onlinecite{1,2,3} and
             \onlinecite{5}, the order--disorder line was considered
             only in the single vortex pinning region.
\end{thebibliography}
\end{document}